\documentclass[amsmath,pra, amssymb,superscriptaddress,twocolumn,notitlepage,nofootinbib]{revtex4-2}
\usepackage[utf8]{inputenc}
\usepackage{amsmath, amssymb, amsthm, physics}
\usepackage[countmax]{subfloat}
\usepackage{booktabs}
\usepackage{bbm}
\usepackage{graphicx}
\usepackage{lineno}
\usepackage{comment}
\usepackage{color}
\usepackage{hyperref}
\usepackage{epstopdf}
\usepackage{dsfont}
\usepackage{multirow}
\usepackage[english]{babel}
\usepackage{graphicx}
\usepackage{float}
\usepackage{lipsum}
\usepackage{curve2e}
\usepackage{hyperref}
\usepackage{xcolor}
\hypersetup{colorlinks=true, linkbordercolor=white, linkcolor=blue, pdfborder={/S/U/W 1}}
\usepackage{braket}
\usepackage{bm}
\usepackage{modiagram}

\begin{document}

\title{Quantum memristor with vacuum--one-photon qubits}

\author{ Simone Di Micco}
\thanks{These two authors contributed equally}
\affiliation{Dipartimento di Fisica - Sapienza Universit\`{a} di Roma, P.le Aldo Moro 5, I-00185 Roma, Italy}

\author{Beatrice Polacchi}
\thanks{These two authors contributed equally}
\affiliation{Dipartimento di Fisica - Sapienza Universit\`{a} di Roma, P.le Aldo Moro 5, I-00185 Roma, Italy}

\author{Taira Giordani}
\email{taira.giordani@uniroma1.it}
\affiliation{Dipartimento di Fisica - Sapienza Universit\`{a} di Roma, P.le Aldo Moro 5, I-00185 Roma, Italy}

\author{Fabio Sciarrino}
\affiliation{Dipartimento di Fisica - Sapienza Universit\`{a} di Roma, P.le Aldo Moro 5, I-00185 Roma, Italy}

\begin{abstract}
Quantum memristors represent a promising interface between quantum and neuromorphic computing, combining the nonlinear, memory-dependent behavior of classical memristors with the properties of quantum states. 
An optical quantum memristor can be realized with a vacuum--one-photon qubit entering a tunable beam splitter whose reflectivity is adapted according to the mean number of photons in the device.
In this work, we report on the experimental implementation of a bulk quantum-optical memristor, working with single-rail coherent superposition states in the Fock basis, generated via a resonantly excited quantum dot single-photon source. We demonstrate that the coherence of the input state is preserved by the quantum memristor. Moreover, our modular platform allows investigating the nonlinear behavior arising from a cascade of two quantum memristors, a building block for larger networks of such devices towards the realization of complex neuromorphic quantum architectures.
\end{abstract}

\maketitle

\textbf{Introduction.} Quantum computing and, in particular, the field of quantum machine learning promise to improve the performances of several tasks and algorithms \cite{Biamonte2017, Wang2024_rev, Cerezo2022}.
In this context, paradigms that mimic the learning process of the human brain with important advantages in terms of energy consumption such as reservoir computing, extreme learning machines, and, more generally, neuromorphic computing, have recently found a generalization in the quantum computing framework 
 \cite{Ghosh2019, Nokkala2021,Innocenti2023, Zambrini_RC,tacchino2020quantum}. Indeed, such schemes require the information to be processed via a fixed random nonlinear high-dimensional network that is particularly suited to be mapped onto quantum channels or circuits \cite{Ghosh2019, Innocenti2023}.
Among the platforms taken into consideration for the realization of quantum machine learning protocols, photons are an interesting candidate, especially because of their compatibility with communication networks, fast information processing, and integrated platforms for quantum computation \cite{lamata2021quantum, Giordani2023,scala2024deterministic,Flamini2018}.
Notably, photonic realizations of quantum machine learning algorithms have already been demonstrated.
In particular, we mention works on proof-of-concept quantum variational circuits \cite{Peruzzo2014,Paesani2017, Wang2017, Santagati2018, Santagati2019, hoch2024variational, facelli2024exactgradientslinearoptics, agresti2024, hoch2024variationalapproachphotonicquantum,Cimini2024}, kernel estimation \cite{hoch2025quantum,yin2024experimentalquantumenhancedkernelsphotonic, Anai_kernel_cv}, neural networks and reservoir computing \cite{ballarini2020polaritonic, 
 Spagnolo2022_memr,Suprano2024, zia2025quantumextremelearningmachines}.

\begin{figure}[tb]
    \centering
    \includegraphics[width=\columnwidth]{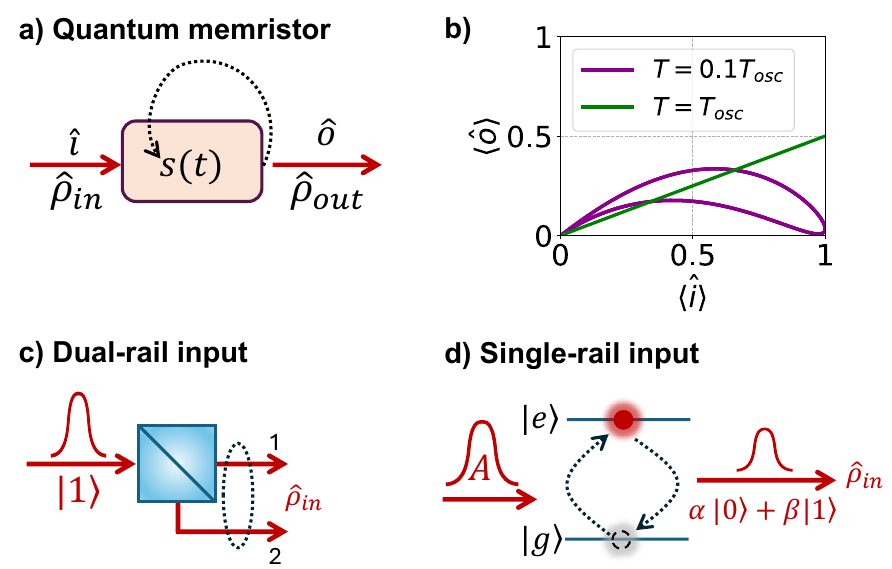}
    \caption{\textbf{Dynamics of a quantum memristor (QM). a)} The dynamics of a QM is defined by its state variable $s(t)$, whose time behavior depends on the input variable $\hat{i}$ and on its previous state $s(t')$, where $t' < t$. \textbf{b)} A QM shows different hysteresis loops depending on the ratio between the input oscillation period $T_{osc}$ and the QM characteristic time $T$. \textbf{c)} Dual-rail encoding of vacuum--one-photon qubits. A photon impinges on a beam-splitter with tunable reflectivity and one of the two output modes is treated as an ancilla \cite{Spagnolo2022_memr}. The resulting state is an entangled state in the form $\hat{\rho}_{in} = \alpha \ket{01}_{12} + \beta \ket{10}_{12}$. \textbf{d)} Here, a laser pulse with pulse area $A$ is used to resonantly excite a semiconductor quantum dot and generate coherent vacuum--one-photon superpositions \cite{loredo2019generation} as input states to the memristor.}
    \label{fig:concept}
\end{figure}

The theoretical proposals for photonic quantum machine learning highlight the introduction of nonlinearities in optical processing as a key element for the effective realization of quantum neural networks \cite{Steinbrecher2019, Spagnolo2023,Stanev2023}. 
One potential method for introducing such nonlinearities is by incorporating measurements with feedback in photonic circuits. In this framework, the Adaptive Boson Sampling schemes \cite{Chabaud2021quantummachine,hoch2025quantum} propose measuring a fraction of photons and applying accordingly adaptive unitary transformations on the unmeasured photons. This variation of the fully linear optical Boson Sampling enables the exploration of the scheme for quantum machine learning tasks, such as kernel estimation \cite{hoch2025quantum}. Alternatively, a way to introduce a memory component in an optical circuit is through a weak measurement with a feedback that depends on the past evolution of the system.
 Examples of such architectures are quantum memristors (QMs) \cite{Sanz2018,lamata2022memristors,pfeiffer2016quantum,salmilehto2017quantum,forsh2024quantum}, i.e., quantum devices that, in the simplest case, are composed of a two-dimensional quantum system, a weak measurement, and feedback system.
Notably, a network of such systems \cite{kumar2021entangled,kumar2022tripartite} may be regarded as a neuromorphic quantum architecture \cite{lamata2024quantum}.
The original idea of a QM originates from the classical memristor model, which is an electrical element whose resistance depends dynamically on the flux of charges in the circuit \cite{Chua_memristor, Chua2018,du2017reservoir}. Classical memristors have also been applied to optics, envisaging the use of active media \cite{Youngblood2023}.
In such a context, the optical QM inserts a nonlinear memory component in the optical processing of single-photon states.
In its original proposal, the QM is embodied by a tunable beam-splitter whose reflectivity is updated in time according to a feedback function that depends on the mean number of photons passing through the device \cite{Sanz2018,Spagnolo2022_memr}.
A recent experimental realization in a linear integrated circuit \cite{Spagnolo2022_memr} adapted such a scheme to dual-rail photonic qubits, at the cost of using an additional ancillary optical mode (see Fig.~\ref{fig:concept}c), {due to the challenge of generating single-rail photon-number superposition states with linear optics \cite{lombardi2002teleportation}}. 
{Nonetheless, recent significant efforts have been devoted to the investigation of photon-number states for quantum information protocols \cite{Polacchi2024,renema2020simulability} and the generation of large entangled states \cite{wein2022photon,vajner2025exploring,santos2023entanglement}.}

In this work, we present the implementation of the original proposal of the optical QM \cite{Sanz2018}, which requires no ancillary modes, since it is based on single-rail vacuum--one-photon qubits \cite{lombardi2002teleportation} (see Fig.~\ref{fig:concept}d).
We generate coherent photon-number qubits through quantum-dot-based single-photon sources \cite{Somaschi2016} that emit mixed or pure states in the vacuum--one-photon encoding \cite{loredo2019generation, Polacchi2024,lombardi2002teleportation}. The QM here presented is represented by a bulk Mach-Zehnder interferometer encoded in the polarization degree of freedom. We show that the so-realized QM preserves the coherence of the quantum state. Furthermore, we test the behavior of two cascaded memristive devices, towards the realization of a network of quantum memristors for quantum machine learning applications.

\begin{figure*}[htb]
    \centering
    \includegraphics[width=1\textwidth]{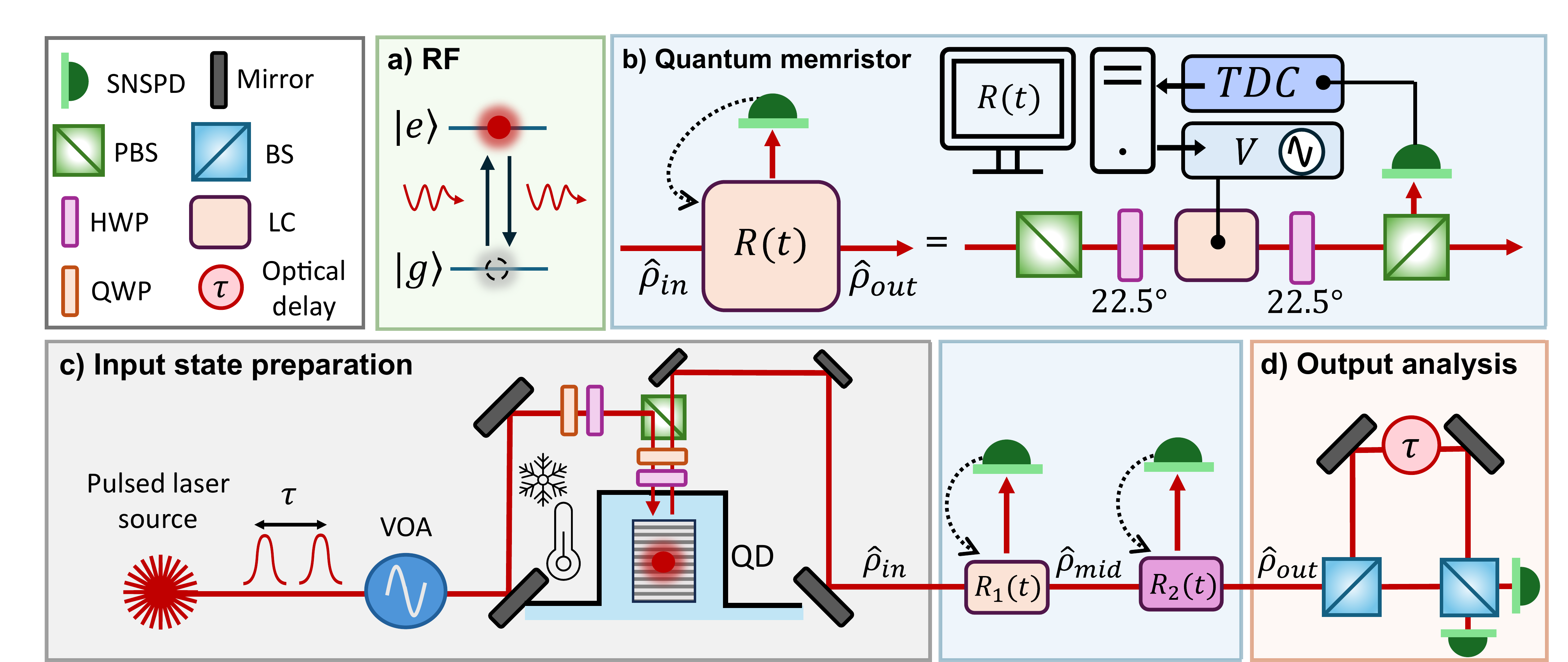}
    \caption{\textbf{Experimental apparatus. a)} Single photons are generated through a neutral and a charged excitonic-based quantum dots (QD), excited through resonance fluorescence (RF) \cite{loredo2019generation}. \textbf{b)} The implemented optical QM consists of two polarizing beam splitters (PBS) with a sequence of two half-waveplates (HWP) rotated by 22.5$^\circ$ and a liquid crystal (LC) in between.
    The feedback loop operates on the phase applied by the LC that controls the overall reflectivity of the QM. \textbf{c)} Input states are prepared through a variable optical attenuator (VOA) to modulate the laser pump power between 0 and $\pi$ as in Eq.~\eqref{eq:in_modulation}. The generated states are collected through a cross-polarization scheme, by using HWPs and quarter-waveplates (QWP). The generated states enter a cascade of two optical QMs, and the output is analyzed in a path-unbalanced MZI. Photons are measured through superconducting nanowire single-photon detectors (SNSPD).}
    \label{fig:exp_app}
\end{figure*}
\textbf{Photonic quantum memristor.} In electronics, a memristor is considered the fourth fundamental electrical component along with the resistor, the capacitor, and the inductor, establishing a relationship between the charge and the flux-linkage \cite{Chua_memristor}. The name \textit{memristor} derives indeed from the contraction of \textit{memory resistor}, as such a device behaves as a non-linear resistor with a memory.
In the most general case, \textit{memristive systems} are defined by the following coupled equations \cite{Chua_1976}:

\begin{equation}
\begin{split}
    o &= f(s, i, t)~i \\
    \dot{s} &= h(s, i, t)
\end{split}
\end{equation}
where $i$ and $o$ indicate the input and output variables, respectively, while $s$ is a state variable representing the memristor state. All variables are assumed to also depend on time $t$.

A similar idea has been applied to quantum systems, and the concept of a QM is depicted in Fig.~\ref{fig:concept}a. A quantum state is sent through a memristive device whose state is updated through a suitable feedback system. The output state is still quantum, and the relationship between the input and the output state is nonlinear, as shown in Fig.~\ref{fig:concept}b. A linear regime is approximated when the feedback time scale is comparable with the input oscillation frequency.

The original proposal of an optical QM \cite{Sanz2018} considers a single-rail vacuum--photon qubit \cite{Polacchi2024,lombardi2002teleportation} as input:
\begin{equation}
    \ket{\psi}_{\text{in}} = \alpha(t) \ket{0} + \beta(t) \ket{1}
    \label{eq:input}
\end{equation}
where the logical qubit states are encoded in the vacuum and one-photon Fock states as $\ket{0}_L \equiv \ket{0} $ and $\ket{1}_L \equiv \ket{1}$.
The mean photon number of the state in Eq.~\eqref{eq:input} varies as $\langle n_{\text{in}} (t) \rangle = |\beta(t)|^2$.
The QM can be then implemented as a tunable beam splitter (BS) whose reflectivity varies through a feedback loop updated as follows \cite{Sanz2018,Spagnolo2022_memr}:
\begin{equation}
\begin{split}
    \langle n_{\text{out}} (t) \rangle &= (1-R(t)) \langle n_{\text{in}} (t) \rangle \\
    \dot{R}(t) &= \langle n_{\text{in}} (t) \rangle - 0.5 
\end{split}
\label{eq:memristor}
\end{equation}
A solution for the update rule defined by Eqs.~\eqref{eq:memristor} is the following \cite{Spagnolo2022_memr}:
\begin{equation}
    R(t) = 0.5 + \frac{1}{T}\int_{t-T}^t(\langle n_{in}(t')\rangle -0.5)dt'
    \label{eq:integral}
\end{equation}
where $T$ is an integration time window used to estimate the mean photon number passing through the QM. Note that necessarily $\langle n_{in}(t')\rangle$ is itself evaluated over a time window, which is the duration of the single exposure $\tau$. Hereafter, we refer to such a quantity as $\langle n_{in}(t')\rangle_\tau$.
As suggested in the original proposal \cite{Sanz2018}, an efficient way to test a photonic quantum memristor is with an always positive and regular input signal, such as the following:
\begin{equation} 
    \langle n_{\text{in}} (t) \rangle = \sin^2 \left({\frac{ \pi}{T_{osc}} t}\right)
    \label{eq:in_modulation}
\end{equation}
where $T_{osc}$ is the oscillation period. The function of the input signal determines the shape and area of the memristor hysteresis loop. With such an input function, the limit regimes for $T \ll T_{osc}$ and $T \approx T_{osc}$ tend respectively to the following behaviors:
\begin{equation}
\begin{split}
    \langle n_{\text{out}} (t) \rangle_{T \ll T_{osc}} &= \langle n_{\text{in}} (t) \rangle - \langle n_{\text{in}} (t) \rangle^2 \\
    \langle n_{\text{out}} (t) \rangle_{T \approx T_{osc}} &= 0.5 \langle n_{\text{in}} (t) \rangle
\end{split}
\label{eq:asymptotic}
\end{equation}
where the second equation shows that the memristor behaves linearly with constant reflectivity (see Fig.~\ref{fig:concept}b).
Different input functions are analyzed in the Supplementary Information.\\

\begin{figure*}[htb]
    \centering
    \includegraphics[width=1\textwidth]{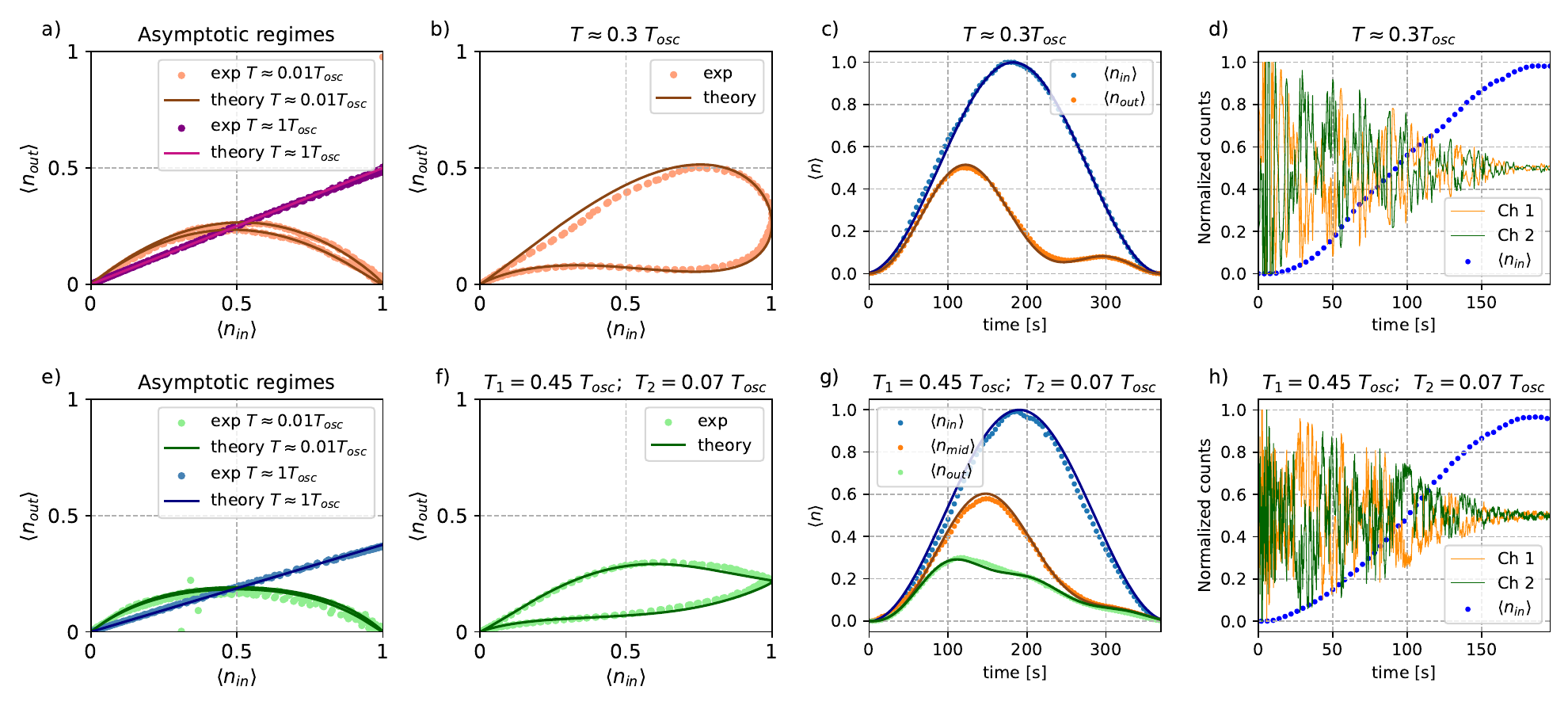}
    \caption{\textbf{Experimental results compared with theoretical simulations.} In panels \textbf{a)} to \textbf{d)} we show the results with a single memristor, while in panels \textbf{e)} to \textbf{h)} the ones with two concatenated memristors. \textbf{a)} Asymptotic regimes described in Eq.~\eqref{eq:asymptotic}, i.e., $T \ll T_{osc}$, and $T \approx T_{osc}$. In the first case we observe a quadratic response, while in the latter we observe a linear behavior. \textbf{b)} Hysteresis loop of a single memristor. The period of the input signal oscillation amounts to $T_{osc} \approx 400$~s. \textbf{c)} Experimental points and expected behavior for $\langle n_{in} \rangle$ and $\langle n_{out} \rangle$ in the time domain, for the time scale shown in panel \textbf{b}. Solid lines represent the theoretical trends. \textbf{e)} Asymptotic regimes for two concatenated memristors with the same integration windows. \textbf{f)} Dynamics of two concatenated memristors with different integration times, as each memristor operates independently. In the figure we called $\langle n_{in} \rangle$ the input state of the first memristor and $\langle n_{out} \rangle$ the output state of the second memristor. \textbf{g)} Experimental points and theoretical expectations for two concatenated memristors in the time domain, for the time scale shown in panel \textbf{f}. Solid lines represent the theoretical trends. With $\langle n_{mid} \rangle$ we refer to the output signal of the first memristor, which is the input signal of the second memristor. \textbf{d-h)} Interference fringes at two single-photon detectors (respectively green and orange) measured by sending the signal outgoing from a single (\textbf{d}) or a concatenation of two (\textbf{h}) memristors to a path-unbalanced MZI. In the background, the correspondent value of the input signal is reported (blue dots). The visibility of the fringes ranges in the interval $[0,1]$ depending on the one-photon component of the input state.}
    \label{fig:results}
\end{figure*}

\textbf{Experimental setup.} 
The experimental setup employed is depicted in Fig.~\ref{fig:exp_app}. It envisages two different \textit{eDelight Quandela} semiconductor quantum dot-based single-photon sources, a charged (QD1) and a neutral (QD2) one \cite{Somaschi2016,loredo2019generation}, excited in a resonance fluorescence (RF) regime (Fig.~\ref{fig:exp_app}a).
Ideally, such sources generate vacuum--one-photon superposition states of the form $\ket{\psi} = \alpha \ket{0} + \beta \ket{1}$, with the one-photon population $|\beta|^2$ that depends on the pump laser power.
In fact, the emitted state is partially mixed and can be modeled as $\hat{\rho}= \mathcal{P} \hat{\rho}_{pure}+(1-\mathcal{P})\hat{\rho}_{mixed}$, where $\hat{\rho}_{pure}=\ketbra{\psi}{\psi}$, $\hat{\rho}_{mixed}= \text{diag}\{|\alpha|^2, |\beta|^2\}$ and $\mathcal{P}$ is known as the conditional purity of the state \cite{loredo2019generation, Polacchi2024}.
The procedure to evaluate the parameter $\mathcal{P}$ is through a self-homodyne detection that is detailed in the Supplementary Information. 
The QMs (Fig.~\ref{fig:exp_app}b) consist of a Mach-Zehnder interferometer (MZI) acting on the polarization degree of freedom. The reflectivity is controlled through a feedback loop that updates the phase shift between the horizontal and vertical polarization inserted by the liquid crystal (LC), according to the measured mean number of photons outgoing the reflected arm of the second polarizing beam splitter (PBS).
The full setup is shown in Fig.~\ref{fig:exp_app}c and further details are given in the Supplementary Information.
The output of the transmitted arm of the two QMs is analyzed firstly by measuring $\langle n_{out} (t) \rangle $ and, in a second experiment, through a self-homodyne measurement in a path-unbalanced MZI to quantify quantum coherence (see Fig.~\ref{fig:exp_app}d and Supplementary Information).

\textbf{Single memristor.} For the characterization of the hysteresis loops of our QMs, we used source QD1. The input signal is modulated as in Eq.~\eqref{eq:in_modulation}.
The choice of the input oscillation period $T_{osc}$ is limited by the feedback system latency ($\tau_{latency} \approx 200$~ms in our case).
Consequently, in order to have $\tau_{latency}\leq 0.05~\tau$ and to observe the regime $T \ll T_{osc}$, we set the input oscillation period to $T_{osc} \simeq 400$~s and the exposure time for the single value of $\langle n_{in}(t) \rangle_\tau$ to $\tau\simeq 4$~s.
With these settings, we were able to retrieve the expected memristive behavior, characterized by the hysteresis cycles observed when plotting $\langle n_{out}\rangle$ vs $\langle n_{in} \rangle $, as depicted in Fig.~\ref{fig:results}a-c. In particular, we report the asymptotic regimes in Fig.~\ref{fig:results}a and the integration time $T = 0.3\, T_{osc}$ in panel b.
Analogous plots at other integration times are reported in the Supplementary Information.
In Fig.~\ref{fig:results}c, instead, we report the temporal shape of the input and output counts, normalized to the single-photon counts at the $\pi$-pulse. The same results have been recovered also by using the source QD2, which emits highly pure superposition states of vacuum--one-photon, unlike source QD1 that generates mostly mixed states. Hence, the purity of the input state does not affect the overall performance of the memristor.

\textbf{Cascaded memristors.} 
In what follows, we take a step further and investigate the evolution of a quantum state passing through two cascaded memristors, realizing a building block for a network of quantum memristors (see Fig.~\ref{fig:exp_app} and Supplementary Information).
Each memristor is updated according to the correspondent reflected mean photon number.
In this configuration, the input state modulates the first reflectivity, and the first output state ($\hat{\rho}_{\text{mid}}$) is used as input to a second memristor. Interesting behaviors appear when the integration times of the two memristors, $T_1$ and $T_2$, are different.
In Fig.~\ref{fig:results}e we report the asymptotic behavior while in
Fig.~\ref{fig:results}f, we show one hysteresis loop of the two cascaded memristors with $T_1 \neq T_2$, where different nonlinearities arise in the two output states, visible, for instance, in the pinched shape at the loop boundaries. In Fig.~\ref{fig:results}g, we show the reconstructed input signal, the signal transmitted by the first memristor, and the one transmitted by the last one. Further measurements of hysteresis loops with different integration times of the two QMs are discussed in the Supplementary Information.

\textbf{Coherence preservation.} To test the memristors' ability to preserve quantum coherence, we used the neutral exciton QD2. Indeed, this source emits states characterized by a high conditional purity $\mathcal{P}_{\text{in}} = 0.95 \pm 0.01$.
The parameter $\mathcal{P}$ can be extrapolated via a self-homodyne detection, by observing the dependence of the fringe visibility on the one-photon population $|\beta|^2$ in single-photon counts recorded when two vacuum--one-photon states interfere in a unbiased beamsplitter, as detailed in the Supplementary Information. In particular, the parameter $\mathcal{P}$ is insensitive to photon losses, i.e. to the presence of incoherent vacuum components in the state  \cite{loredo2019generation, Polacchi2024}. Since each QM introduces only incoherent vacuum terms such as $R(t)\ket{0}\bra{0}$ \cite{Spagnolo2022_memr}, we expect that it does not affect the conditional purity $\mathcal{P}$ of the output states. 
The coherence in the state can be assessed by observing Fig.~\ref{fig:results}d and Fig.~\ref{fig:results}h, where we compare the single-photon counts normalized to the sum of the two output channels of the MZI, passing through one or two memristors, respectively.
The presence of fringes in the output channels demonstrates that the quantum coherence is preserved after each memristor.
The coherence is quantified by measuring the same parameter $\mathcal{P}$ of the output state after the operation of the two QMs, through self-homodyne measurements in a path-unbalanced MZI previously described. It amounts to $\mathcal{P}_{\text{single}} = 0.99 \pm 0.01$ for the state outgoing the first memristor and $\mathcal{P}_{\text{double}} = 0.987 \pm 0.003$ for the state outgoing the two-memristor cascade. Both values are compatible with the purity of the input state within less than three standard deviations.

\textbf{Discussions.}
We provided the experimental demonstration of the fundamental ingredients to implement the original proposal of an all-optical QM by using quantum dot-based single-photon sources to generate superpositions of vacuum and one-photon states on a single spatial mode.
The QM here considered does not affect the conditional purity of the output state, which is fully preserved. Moreover, the proposed vacuum--one-photon QM is versatile and modular, as we show by concatenating two of them in a two-node network, and can be integrated into different network topologies.
Further characteristics of the photonic QM here proposed are analyzed in the Supplementary Information, such as the memristive behavior for different input functions, the effect of losses on the observed coherence, and the topology of the network employed.
The photon-number encoding here used has the advantage of requiring no ancillary modes and was recently shown compatible with the generation of photon-number entanglement \cite{wein2022photon, vajner2025exploring,santos2023entanglement} and quantum information protocols \cite{Polacchi2024,renema2020simulability}. On the other hand, the main bottleneck in its usage resides in its high sensitivity to optical losses.
However, significant efforts are being directed toward improving the extraction efficiency of single-photon sources based on quantum dots \cite{Wang2019Towards,margaria2024efficient} and single-photon detection \cite{natarajan2012superconducting}, which encourages the investigation of scalable platforms supporting this encoding.
Indeed, QMs are suitable to scale up quantum neural networks and quantum neuromorphic computing architectures, even compatibly with near-term quantum devices \cite{lamata2024quantum}.

\begin{acknowledgments}
We acknowledge support from the ERC Advanced Grant QU-BOSS (QUantum advantage via nonlinear BOSon Sampling, grant agreement no. 884676), from the European Union’s Horizon Europe research and innovation program under EPIQUE Project (Grant Agreement No. 101135288), and from PNRR MUR project
PE0000023-NQSTI (National Quantum Science and Technology Institute, Spoke 4).
\end{acknowledgments}

\section*{Competing Interests}
The authors declare no competing interests.

\section*{Data availability}
The data that support the findings of this study are available from the corresponding author upon reasonable request.


%

\end{document}


\title{Supplementary Information: Quantum memristor with vacuum--one-photon qubits}

\author{Simone Di Micco}
\thanks{These two authors contributed equally}
\affiliation{Dipartimento di Fisica - Sapienza Universit\`{a} di Roma, P.le Aldo Moro 5, I-00185 Roma, Italy}

\author{Beatrice Polacchi}
\thanks{These two authors contributed equally}
\affiliation{Dipartimento di Fisica - Sapienza Universit\`{a} di Roma, P.le Aldo Moro 5, I-00185 Roma, Italy}

\author{Taira Giordani}
\email{taira.giordani@uniroma1.it}
\affiliation{Dipartimento di Fisica - Sapienza Universit\`{a} di Roma, P.le Aldo Moro 5, I-00185 Roma, Italy}

\author{Fabio Sciarrino}
\affiliation{Dipartimento di Fisica - Sapienza Universit\`{a} di Roma, P.le Aldo Moro 5, I-00185 Roma, Italy}

\maketitle

\section{Details about the experimental setup}

The experimental platform reported in Fig.~2 of the main text envisages two semiconductor quantum dot-based single-photon sources, a charged (QD1) and a neutral (QD2) one \cite{loredo2019generation}. In particular, we employ commercial InGas quantum dots embedded in electrically controlled micro-cavity (\textit{eDelight Quandela} sources) \cite{Somaschi2016}.
A neutral semiconductor quantum dot emits photons due to the recombination of a neutral exciton, i.e., a bound state of an electron and a hole, while a charged one has an additional electron or hole, forming charged excitons. 
Both QDs are optically excited by a resonant pulsed laser that matches the cavity wavelength, in detail, $\lambda_1 = 927.5$~nm and $\lambda_2 = 928.05$~nm (Fig.~2a of the main text). In Supplementary Fig. \ref{fig:RF_source} we report the observation of the typical Rabi oscillation for the two sources that demonstrates the ability to control the vacuum population in the state.
Ideally, such an excitation scheme would enable the generation of pure vacuum--one-photon superposition states of the form $\ket{\psi} = \alpha \ket{0} + \beta \ket{1}$, with the one-photon population $|\beta|^2$ depending on the pump laser power.
In a realistic setup, the emitted state is a mixed state $\hat{\rho}= \mathcal{P} \hat{\rho}_{pure}+(1-\mathcal{P})\hat{\rho}_{mixed}$, where $\hat{\rho}_{pure}=\ketbra{\psi}{\psi}$ and $\hat{\rho}_{mixed}= \text{diag}\{|\alpha|^2, |\beta|^2\}$.
The one-photon population $|\beta|^2$ is linked to the fringe visibility $\nu$ measured via self-homodyne detection in a path-unbalanced MZI and to the conditional purity parameter $\mathcal{P}$ according to the following formula \cite{loredo2019generation,Polacchi2024}:
\begin{equation} \label{eq:visibility_vs_lambda}
    \nu = \mathcal{P}^2 (1-|\beta|^2) \sqrt{V_{\text{HOM}}}
\end{equation}
Hence, by measuring the fringe visibility for different values of the pump power ranging from 0 to $\pi$, one can evaluate the purity parameter $\mathcal{P}$. The linear fit performed to extract this value for QD2 is shown in Suppl. Fig.~\ref{fig:RF_source}b. The purity of QD1 is $\sim 0$ that means it emits $\rho_{mixed}$ states.
We employ QD2 to demonstrate that the quantum memristor preserves the coherence of the input quantum state, while QD1 is used to demonstrate the memristive behavior of the device.

Source QD1 emits photons at the count rate of $5$~MHz when excited by a pulsed laser with a repetition rate of $79$~MHz.
The typical single-photon purity of this source, measured as the  second-order correlation function, is $g^2_{\text{1}}(0)\sim 3.4\%$, and the indistinguishability of the photons is around $V_{\text{HOM,1}}\sim 80\%$.

Source QD2 is a neutral exciton has a typical single-count rate of $7$~MHz, $g^2_{\text{2}}(0) \sim 0.65 \%$, photons indistinguishability $V_{\text{HOM,2}}\sim 91.5 \%$ and a high conditional purity of vacuum--one photon qubits $\mathcal{P} \sim 95\%$ (see Suppl. Fig.~\ref{fig:RF_source}).

The QM consists of a Mach-Zehnder interferometer (MZI) encoded in the polarization degree of freedom. 
More in detail, as visible in Fig.~2b of the main text, this consists of a polarizing beam splitter (PBS) to prepare a horizontal state, a half-waveplate (HWP) rotated at $22.5^{\circ}$, a liquid crystal with the optic axis parallel to the horizontal polarization, and finally another HWP and a PBS to measure the horizontal and vertical component of the state. Overall, this system acts as a BS with tunable reflectivity, depending on the external voltage applied to the LC, which fully controls the inserted phase shift and, consequently, the fraction of horizontal and vertical components exiting the PBS.
The feedback loop works as follows. A time-to-digital converter (TDC) interfaced with a superconducting nanowire detector (SNSPD) keeps track of the mean number of photons reflected by the MZI. Then, a routine computes the new reflectivity $R(t)$ according to Eq.~(5) of the main text and updates the voltage applied to the LC.
We program a variable optical attenuator (VOA) to modulate in time the power of the excitation pulses between 0 and $\pi$, at frequency $\frac{1}{T_{osc}}$, in such a way that the quantum dot source generates single-photon states with a mean photon number varying as in Eq.~(6) of the main text. The so-generated states enter two cascaded QMs which are updated according to the reflected mean photon number they record individually.
The output of the transmitted arm of the QM is analyzed through self-homodyne measurements in a path-unbalanced MZI (see Fig.~2d of the main text) where the optical delay $\tau$ applied to one arm is equal to the pulse separation of the laser pump.

\begin{figure}[t]
    \centering
    \includegraphics[width=0.45\linewidth]{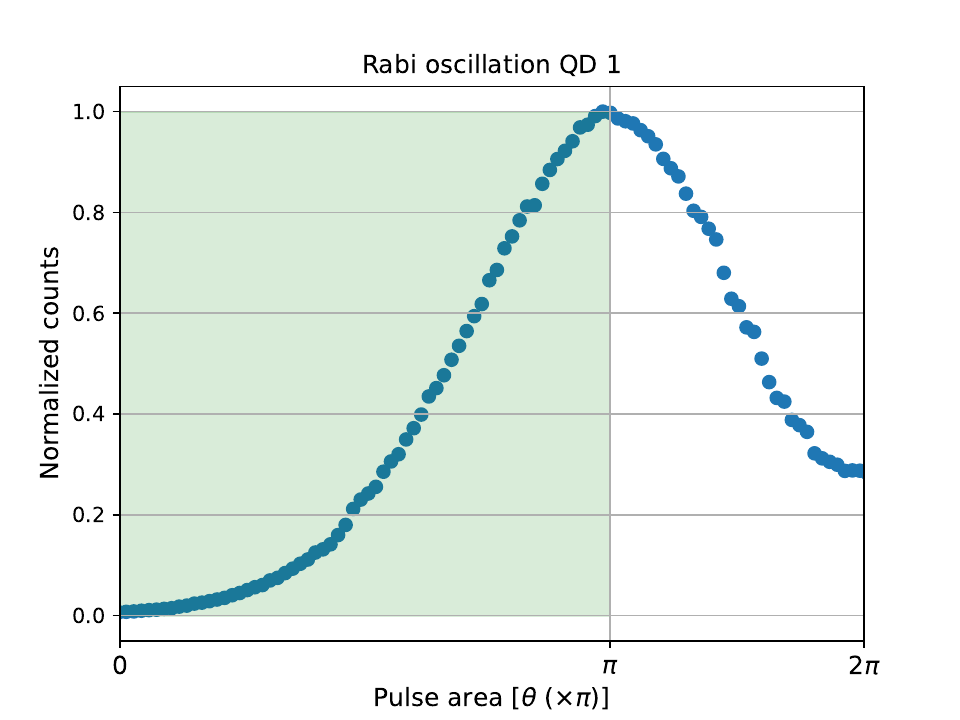}
    \includegraphics[width=0.45\linewidth]{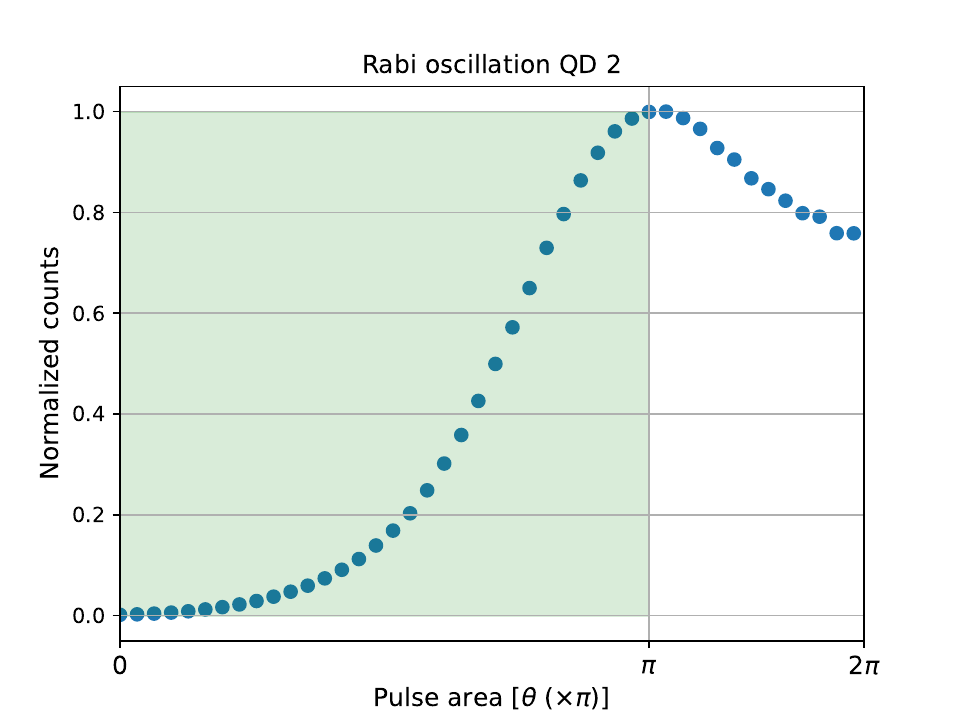}
    \includegraphics[width=0.45\linewidth]{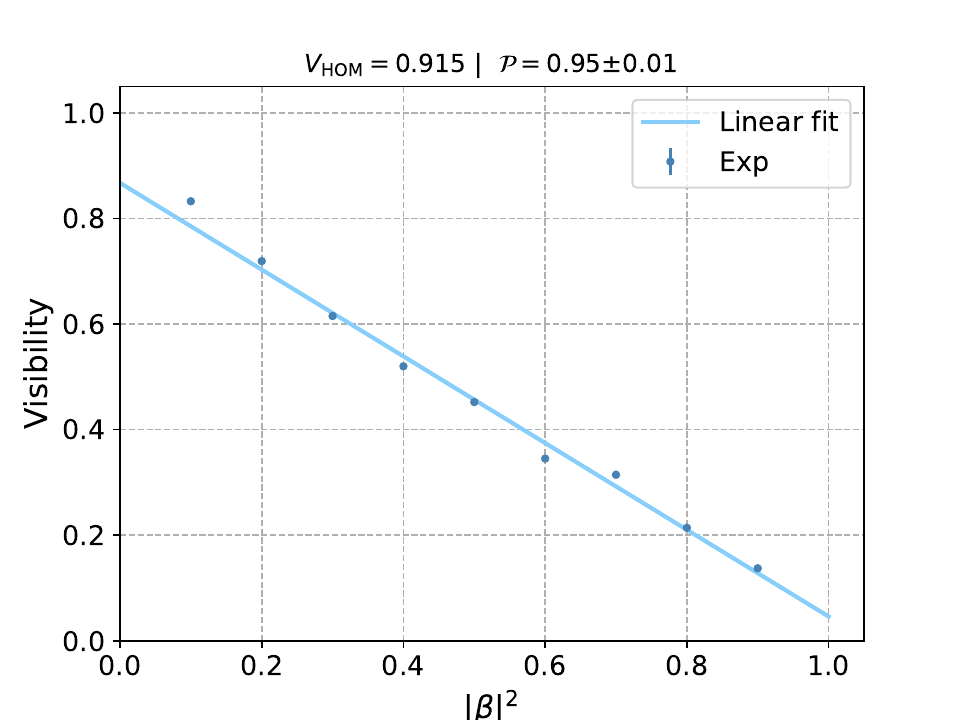}
    \caption{\textbf{Rabi oscillation.} In the two upper figures we report the emitted single-count rates of QD1 (left) and QD2 (right) normalized to their maximum as a function of increasing excitation power. Resonant excitation yields a Rabi oscillation in the two-level artificial atom represented by the quantum dots. With this plot, we show that we are able to generate genuine vacuum--one-photon superposition states when increasing the pump power from 0 to the $\pi$-pulse (green area). \textbf{Purity of QD2.} In the lower figure we characterized the fringe visibility of the state generated by QD2 through self-homodyne measurements in a path-unbalanced MZI, as a function of the one-photon population.}
    \label{fig:RF_source}
\end{figure}

\section{Experimental imperfections and conditional purity of the output state}

\begin{figure}[ht!]
    \centering
    \includegraphics[width=0.4\linewidth]{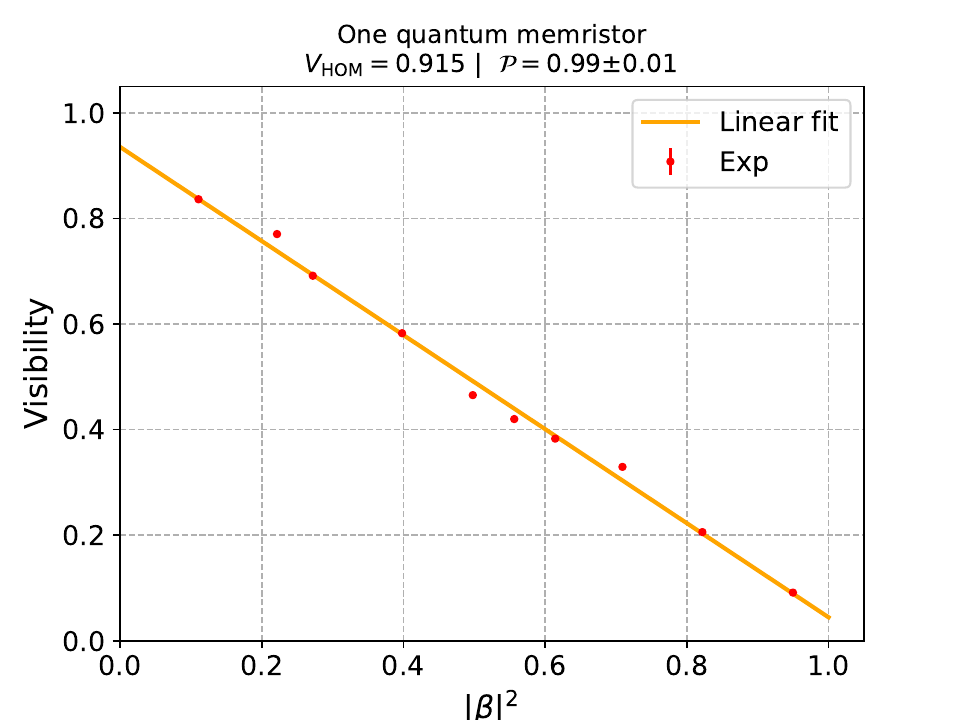}
    \includegraphics[width=0.4\linewidth]{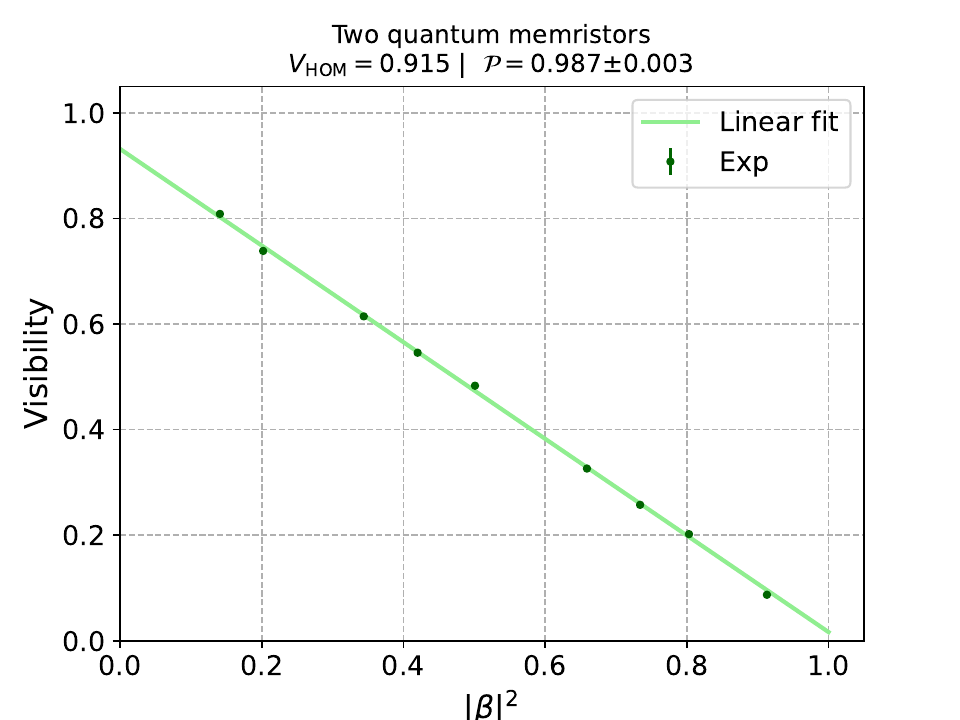}
    \caption{\textbf{Purity of the state outgoing the memristor.} Linear fit of the fringe visibility as a function of the one-photon population of the state outgoing one memristor (left) and two cascaded memristors (right). }
    \label{fig:purity_one_memr}
\end{figure}

In the following calculations, we show that the interference fringes visibility recorded in a self-homodyne detection for a state outgoing a QM does not depend on the reflectivity caused by the quantum memristor and, therefore, the conditional purity is unchanged. By conditional purity, we mean the purity measured upon the detection of single photons. Furthermore, we consider the effects of experimental imperfections such as losses and imperfect photon indistinguishability in this model.

Let us consider two vacuum--one-photon qubits generated at two consecutive excitations at time $t=0$ and $t=\tau$:
\begin{equation}
\ket{\psi}_{in} = (\alpha + \beta e^{i\phi} a_1^{\dagger})_{\tau} \otimes (\alpha + \beta  a_1^{\dagger})_0 \ket{0}
\end{equation}

where $\beta$ is real, for simplicity. After the quantum memristor:
\begin{equation}
\ket{\psi} = (\alpha + \beta e^{i\phi} (\sqrt{T }a_1^{\dagger} + \sqrt{R}a_3^{\dagger}) )_{\tau} \otimes (\alpha + \beta (\sqrt{T }a_1^{\dagger} + \sqrt{R}a_3^{\dagger}))_0 \ket{0}
\label{eq:after_mem}
\end{equation}

where we have defined the reflectivity and trasmissivity of the QM as $R=R(t)$ and $T = 1-R(t)$ for brevity. After the first BS of the MZI:
\begin{equation}
\ket{\psi} = \left(\alpha + \beta e^{i\phi} \left(\sqrt{\frac{T}{2}}(a_1^{\dagger} + a_2^{\dagger}) + \sqrt{R}a_3^{\dagger}\right) \right)_{\tau} \otimes \left(\alpha + \beta  \left(\sqrt{\frac{T}{2}}(a_1^{\dagger} + a_2^{\dagger}) + \sqrt{R}a_3^{\dagger}\right) \right)_0 \ket{0}
\end{equation}

After $\tau$ delay on mode 2:
\begin{equation}
\ket{\psi} = \left(\alpha + \beta e^{i\phi} \left(\sqrt{\frac{T}{2} }(a_{1, \tau}^{\dagger} + a_{2, 2 \tau}^{\dagger}) + \sqrt{R} a_{3,\tau}^{\dagger}\right) \right) \otimes \left(\alpha + \beta \left(\sqrt{\frac{T}{2} }(a_{1,0}^{\dagger} + a_{2,\tau}^{\dagger}) + \sqrt{R}a_{3,0}^{\dagger}\right)\right) \ket{0}
\end{equation}

After second BS of the MZI:
\begin{equation}
\begin{split}
\ket{\psi} &= \left(\alpha + \beta \left(\frac{\sqrt{T}}{2}  (a_{1,0}^{\dagger}  + a_{2,0}^{\dagger} + a_{2,\tau}^{\dagger} - a_{1,\tau}^{\dagger}) + \sqrt{R}a_{3,0}^{\dagger}\right)\right) \ket{0} \otimes \\
&\otimes \left(\alpha + \beta  e^{i \phi} \left(\frac{\sqrt{T}}{2} (a_{1, \tau}^{\dagger} + a_{2, \tau}^{\dagger} + a_{2, 2 \tau}^{\dagger} - a_{1, 2 \tau}^{\dagger}) + \sqrt{R} a_{3,\tau}^{\dagger}\right) \right)
\end{split}
\label{eq:after_MZI}
\end{equation}

When using threshold detectors, the probability of detecting a photon in mode 1 at time $\tau$ is given by: 
\begin{equation}
    P({1,\tau}) = |\alpha|^2|\beta|^2 T \sin(\phi/2)^2 + \frac{|\beta|^4 R T}{2} + \frac{3}{8}|\beta|^4 T^2
\end{equation}
The visibility $\nu$ of such a trace is:
\begin{equation}
    \nu = \frac{P({1,\tau})^{Max}- P({1,\tau})^{Min}}{P({1,\tau})^{Max}+ P({1,\tau})^{Min}}= \frac{4 - 4|\beta|^2}{4 - |\beta|^2T} 
    \label{eq:vis_single}
\end{equation}
However, when considering an overall photon collection efficiency $\eta$, the probability of detecting one photon at time $\tau$ to the fringes becomes:
\begin{equation}
    P({1,\tau}) = |\alpha|^2|\beta|^2 T \sin(\phi/2)^2\eta + \frac{|\beta|^4 R T}{2} \eta + \frac{1}{4}|\beta|^4 T^2\eta + \frac{1}{8}|\beta|^4 T^2\eta(2 - \eta)
\end{equation}
where the last term refers to the bunching contribution. When $\eta \rightarrow 0$, as in our case (photon extraction efficiency $\approx 0.15$ and photon detection $\approx 0.85$), the visibility tends to the linear function $\nu = 1-|\beta|^2$. When also considering the conditional purity of the quantum states emitted by the source and the photon partial distinguishability, estimated through the HOM visibility, the expected fringe visibility amounts to $\nu =  \mathcal{P}^2 (1-|\beta|^2) \sqrt{V_{\text{HOM}}}$, as in the case without memristor. The last two sources of imperfections can be included as follows. The partial indistinguishability is simulated by modeling the creation operator as $a_{1}^{\dagger} \rightarrow \sqrt{x} a_{1}^{\dagger}+\sqrt{1-x} a_{1, \perp}^{\dagger}$, where $x =  \sqrt{V_{HOM}}$ and $a_{1, \perp}^{\dagger}$ corresponds to the creation in the same path of one photon in a perfect orthogonal state in the other internal degrees of freedom, for example, polarization or time arrival. The imperfect conditional purity reflects the fact that the emitted state is a mixed one like $\hat{\rho}= \mathcal{P} \hat{\rho}_{pure}+(1-\mathcal{P})\hat{\rho}_{mixed}$, where $\hat{\rho}_{pure}=\ketbra{\psi}{\psi}$ and $\hat{\rho}_{mixed}= \text{diag}\{|\alpha|^2, |\beta|^2\}$. 
Then, the conditional purity of the output quantum states after the QM can be therefore extracted from the linear fits shown in Suppl. Fig.~\ref{fig:purity_one_memr}. 

The same arguments hold for the two concatenated memristors. This time the state in eq. \eqref{eq:after_mem} becomes:
\begin{equation}
\ket{\psi} = (\alpha + \beta e^{i\phi} (\sqrt{T _1 T_2}a_1^{\dagger} + \sqrt{R_1}a_3^{\dagger} + \sqrt{T_1R_2}a_4^{\dagger}) )_{\tau} \otimes (\alpha + \beta  (\sqrt{T _1 T_2}a_1^{\dagger} + \sqrt{R_1}a_3^{\dagger} + \sqrt{T_1R_2}a_4^{\dagger}) )_0 \ket{0}
\label{eq:after_double_mem}
\end{equation}
where $a_4^{\dagger}$ is the feedback channel of the second memristor.
After the evolution in the MZI, the probability of detecting one photon at time $\tau$ at one of the two outputs is

\begin{equation}
    P({1,\tau})_{double} = |\alpha|^2|\beta|^2 T_1 T_2 \sin(\phi/2)^2 + \frac{|\beta|^4 R_1 T_1 T_2}{2} + \frac{|\beta|^4 R_2 T_1^2 T_2}{2}+ \frac{3}{8}|\beta|^4 T_1^2 T_2^2.
\end{equation}

The visibility of the fringes will be

\begin{equation}
    \nu_{double} = \frac{4 - 4|\beta|^2}{4 - |\beta|^2T_1T_2} 
    \label{eq:vis_double}
\end{equation}

The dependence on the QMs' reflectivity vanishes again in the condition of high losses as it was in the case of a single memristor. Then, for $\eta \rightarrow 0$ that reflects the experimental conditions and for imperfect photon indistinguishability and conditional purity $\mathcal{P}$, we expect $\nu_{double} \rightarrow \mathcal{P}^2(1-|\beta|^2) \sqrt{V_{HOM}}$. Such a behavior is confirmed in the measurements reported in the right panel of Suppl. Fig. \ref{fig:purity_one_memr}. The value of $\mathcal{P}$ estimated from the linear fit is in accordance with those observed with a single-memristor and without any memristor.

\begin{figure}[t]
    \centering
    \includegraphics[width=0.9\linewidth]{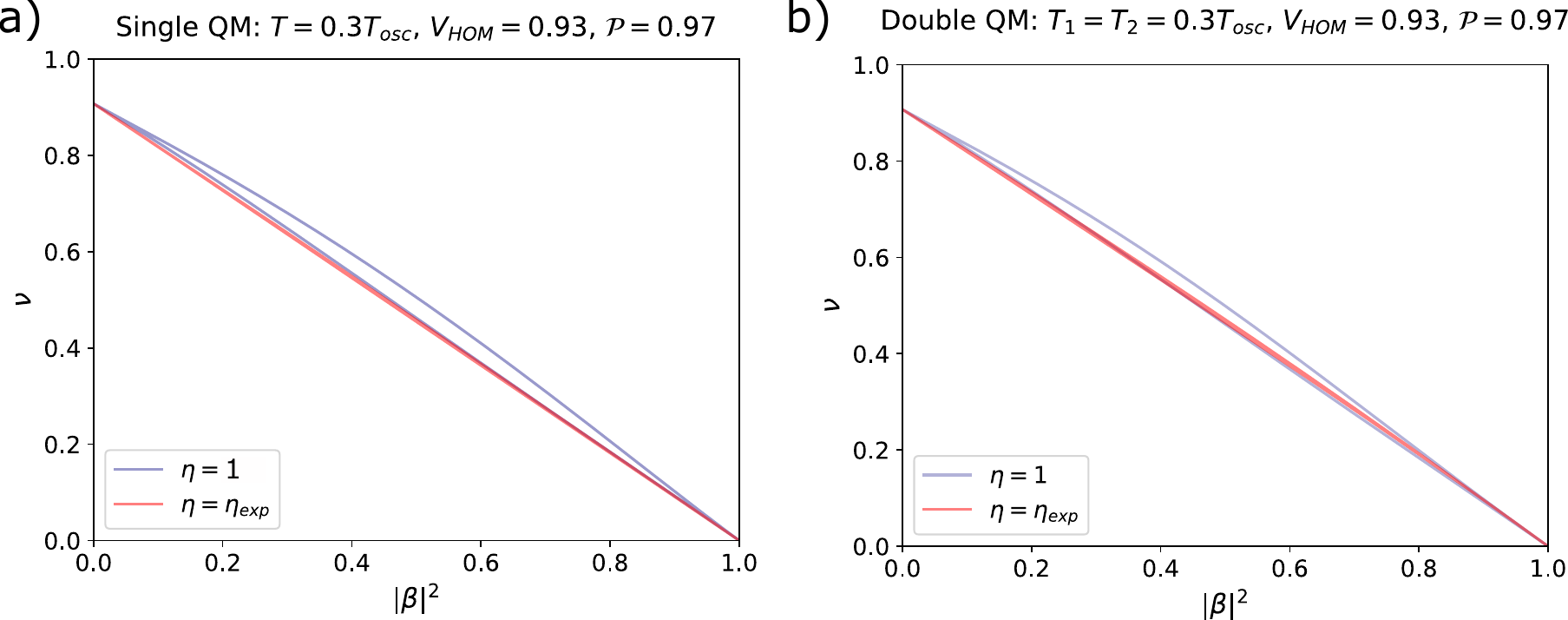}
    \caption{\textbf{Visibility of interference fringes after quantum memristors in imperfect apparatuses.} a) single QM case in the ideal lossless scenario $\eta\rightarrow1$ (blue) and in the experimental losses condition (red). b) Same study for the two concatenated QMs. All curves take into account the typical photons indistinguishability and conditional purity of the states emitted by QD2.}
    \label{fig:losses}
\end{figure}

Note that even in a lossless scenario, the concatenation of $n$ QMs would lead to an average total trasmissivity of the setup $T=\prod_{i=1}^n  T_i \rightarrow 0$ that brings the fringes' visibility to behave as in a lossy scenario. This is analyzed in the simulation carried out in Suppl. Fig. \ref{fig:losses}. Moreover, Eqs.~\eqref{eq:vis_single},\eqref{eq:vis_double} show that in lossless conditions the visibility displays a small hysteresis loop, that vanishes in a lossy apparatus as the experimental one or in the presence of several concatenated QMs. This can be observed by the curves for $\eta \rightarrow1$: the one of the two concatenated memristors display a smaller hysteresis loop as expected by the calculations carried out in this section. In fact, as also discussed in the main text and in \cite{Spagnolo2022_memr}, the effect of the QM on $\nu$ when tracing on the feedback channel is analogous to the one of a loss. 

\section{Building block for a network of quantum memristors}

\begin{figure}[h!]
    \centering
    \includegraphics[width=0.55\linewidth]{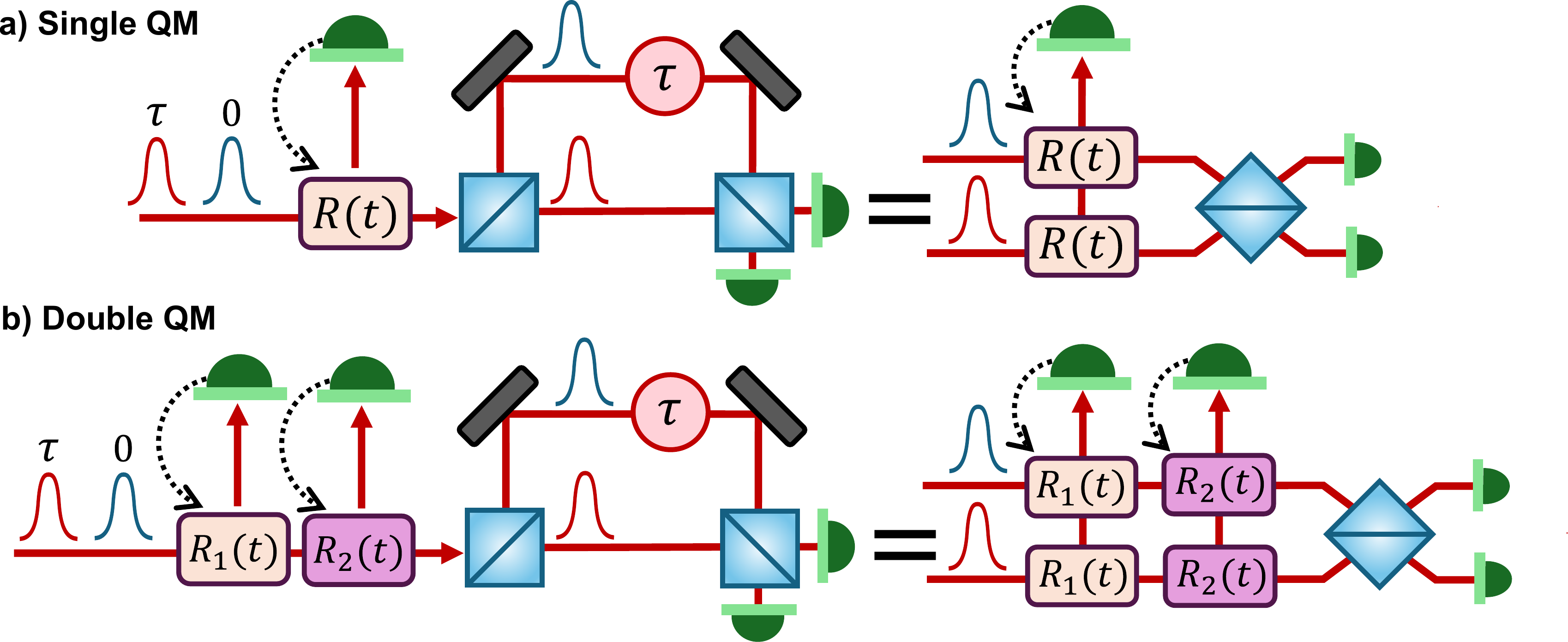}
    \caption{\textbf{Equivalence between cascaded memristors in the time-bin encoding and parallel memristors in the spatial encoding in the single a) and double b) case.} Sending two consecutive states through a chain of one (\textbf{a)}) or two QMs (\textbf{b)}) and then analyzing them through interference on a BS is equivalent to let them pass independently through two parallel copies of the same memristor series and then let them interfere on a BS.}
    \label{fig:schemini}
\end{figure}

We show here the equivalence between the series of two memristors presented in the main text and a configuration with parallel memristors. A detailed scheme is shown in Suppl. Fig.~\ref{fig:schemini}.
In detail, the time-bin encoding of two consecutive pulses that pass through a chain of memristors and then interfere in a MZI can also be seen as the two pulses traveled along parallel rails and interfered in a beam-splitter. However, we note that, in our setup, this equivalence is probabilistic and conditioned on correct routing at the first BS of the MZI.
Since the configuration here proposed is equivalent to have both cascaded and parallel memristors, it can be seen as a building block for a network of quantum memristors.

\section{Quantum memristor effect at different integration times}

\begin{figure}[h!]
    \centering
    \includegraphics[width=0.8\textwidth]{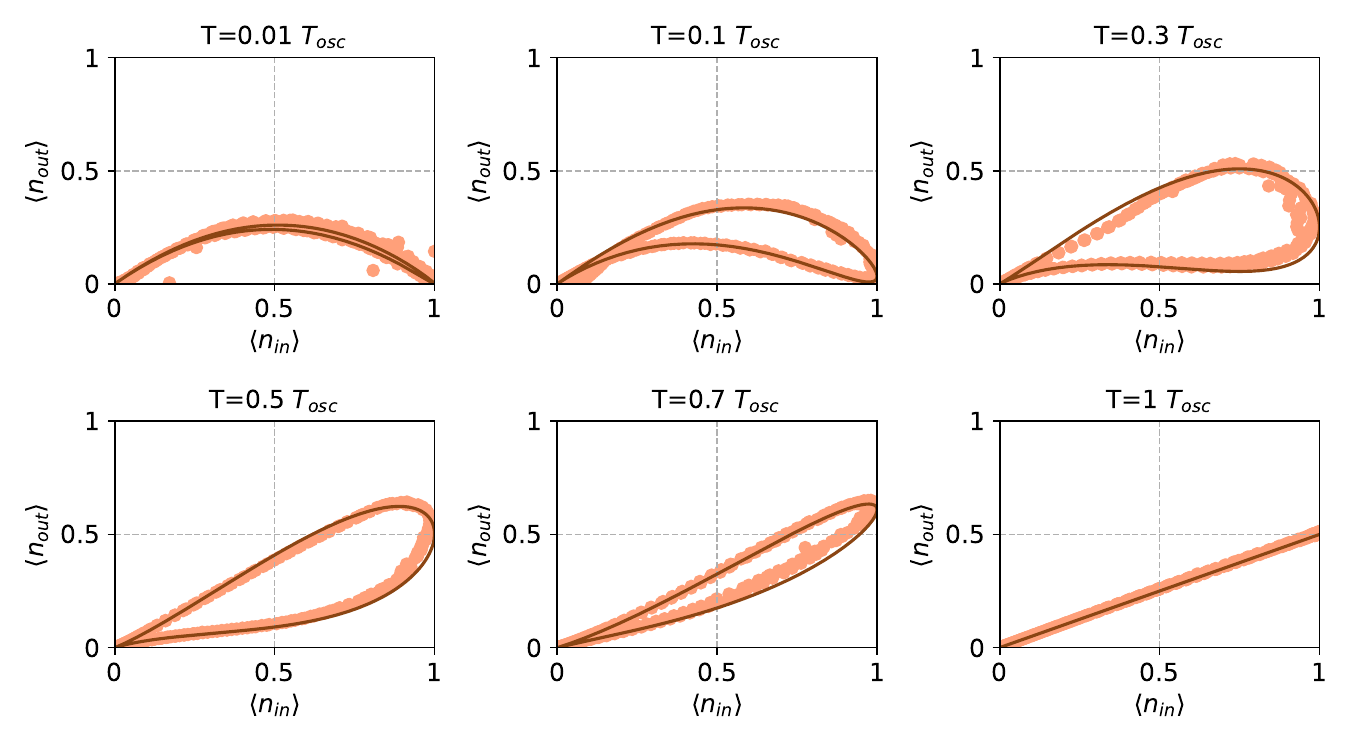}
\caption{\textbf{Dynamics of a memristor.} We show the experimental hysteresis loops when plotting the average number of photons at the transfer mode of the memristor vs the average number of photons at the input mode. The experimental data are compared with the theoretical expectations, represented with smooth lines, when the input signal is modulated through the function in Eq.~[5] in the main text, and the reflectivity is updated using the Eq.~[4] in the main text. When the integration time $T \ll T_{osc}$, the quantum memristive behavior approximates $\langle n_{out}\rangle \approx \langle n_{in}\rangle - \langle n_{in}\rangle^2$ while, when the integration time is comparable with the input signal period $T \approx T_{osc}$, the behavior becomes linear and equal to $\langle n_{out}\rangle \approx 0.5 \langle n_{in}\rangle $. The period of the input signal oscillation is $T_{osc} \approx 400$~s, while the time windows for the single exposure is $\tau=4$~s.}
    \label{fig:cicli}
\end{figure}

\begin{figure}[h!]
    \centering
    \includegraphics[width=0.8\textwidth]{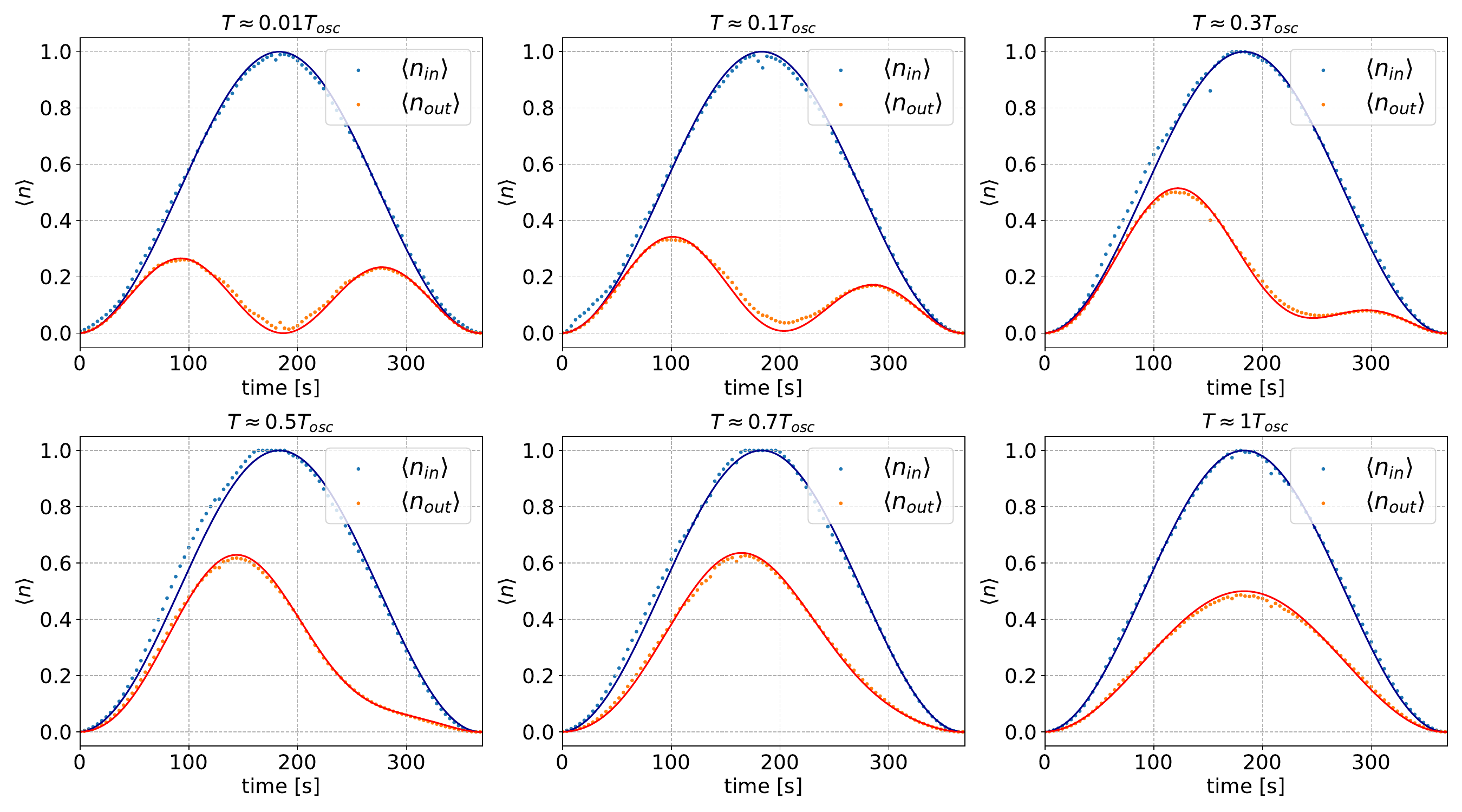}
    \caption{\textbf{Temporal shapes.} In the figure we show the average number of photons in the time domain for both the input and transmitted mode of the memristor for different integration times T. The experimental data (dots) are compared with the theoretical expectations, which are depicted with smooth lines.}
    \label{fig:temporal_shapes}
\end{figure}

\begin{figure}[h!]
    \centering
    \includegraphics[width=0.8\textwidth]{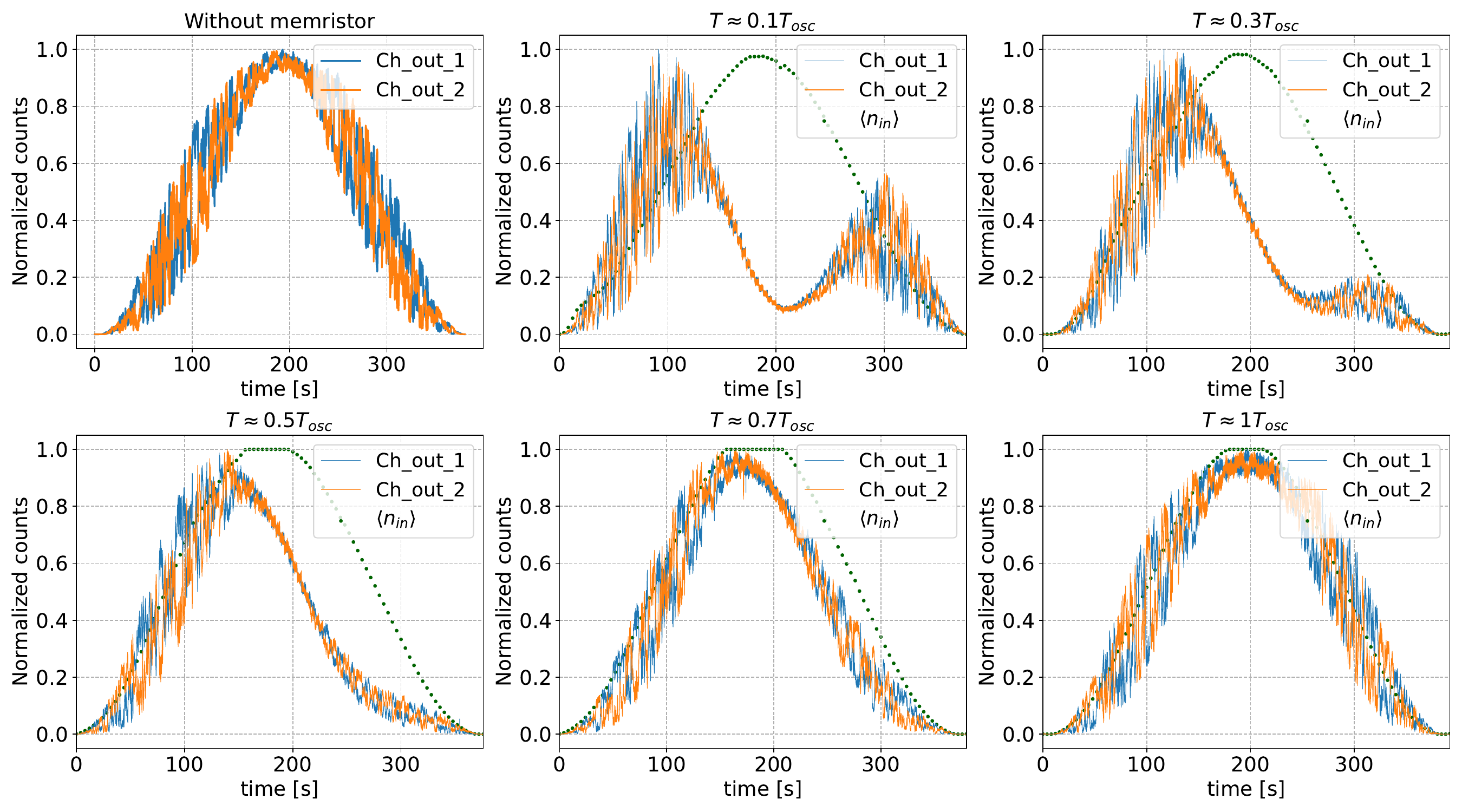}
    \caption{\textbf{Interference fringes in the time domain.} In the figure we show the interference fringes obtained by making interfere two subsequent state outing from the transfer mode of the memristor in a self-homodyne scheme. In order to make them interfere we sent them to a path-unbalanced MZI. The smooth lines represent the signals measured at the two output channels of the MZI, while the dots represent the input signal reconstructed by using the reflected component of the memristor and the current value of reflectivity and the value of the transmitted component of the memristor. For the reconstruction of the input signal it was imposed a control in such a way that the value does not surpass the unit. In case it did, $\langle n_{in} \rangle$ was set to 1.}
    \label{fig:fringes}
\end{figure}

\begin{figure}[h!]
    \centering
    \includegraphics[width=0.8\textwidth]{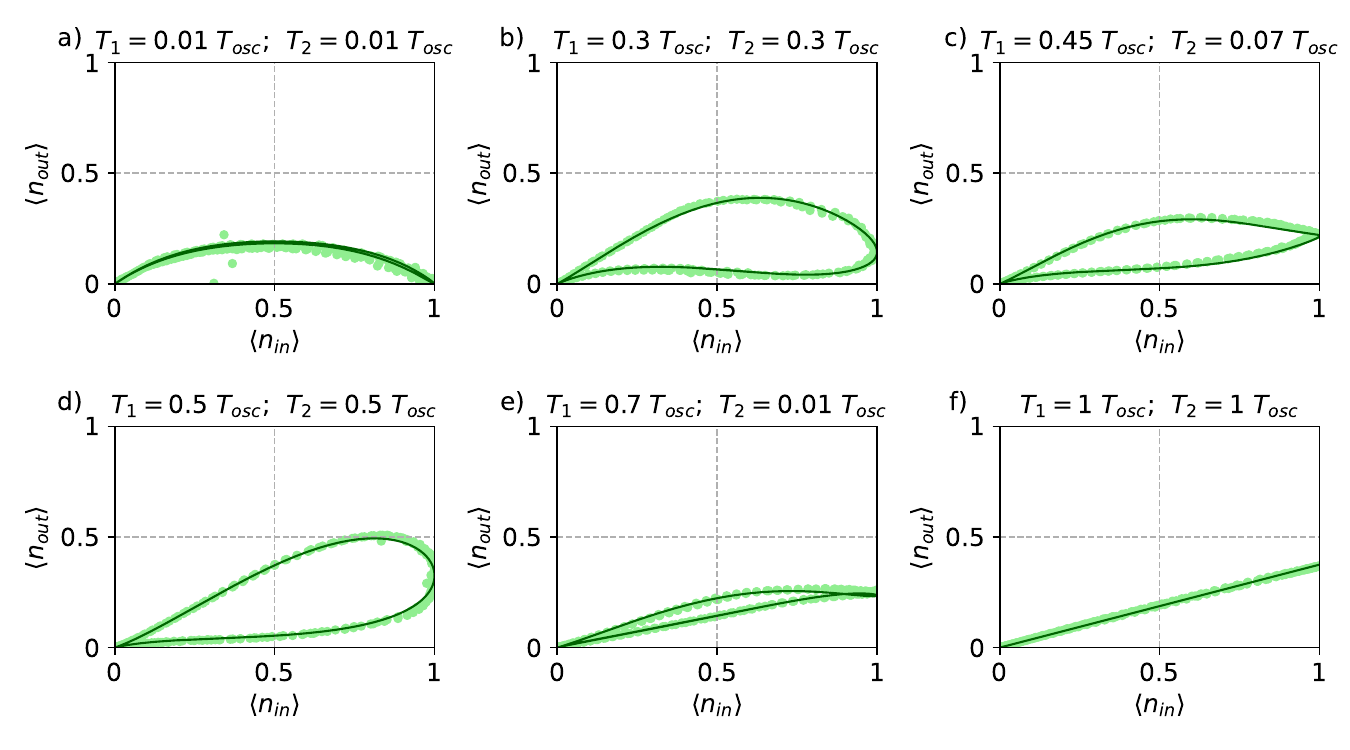}
    \caption{\textbf{Dynamics of concatenated memristors.} We show the experimental hysteresis loops of two concatenated QMs. As the memristors operate independently, we could feature them with the same integration time or with different integration times. The experimental data (dots) are compared with the theoretical expectations (smooth line). The period of the input signal oscillation is kept at the value $T_{osc} \approx 400$~s.}
    \label{fig:concatenated_equal_memristors}
\end{figure}

\begin{figure}[h!]
    \centering
    \includegraphics[width=0.8\textwidth]{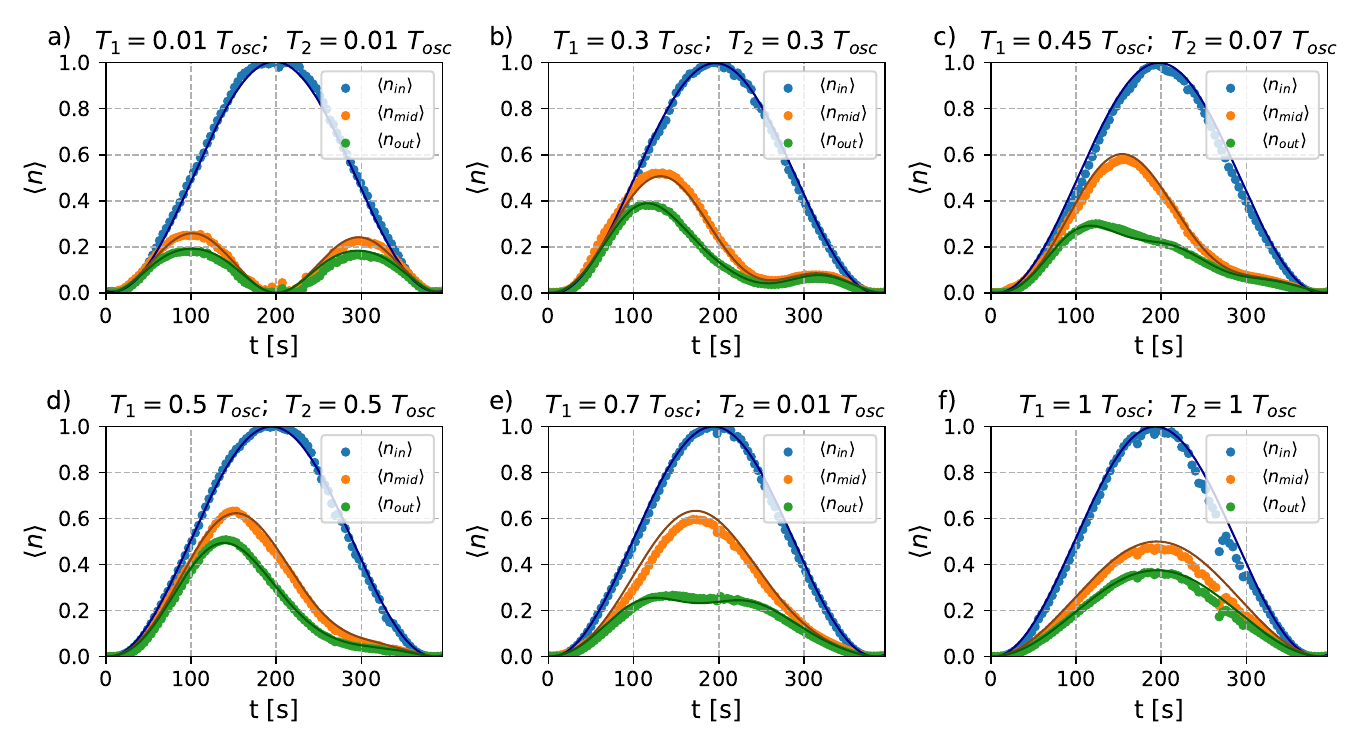}
    \caption{\textbf{Temporal shapes.} In the figure we show the average number of photons in the time domain for the input mode of the first memristor ($\langle n_{in} \rangle$), the output mode of the first memristor ($\langle n_{mid} \rangle$), and the transmitted mode of the second memristor ($\langle n_{out} \rangle$) for different integration times $T_1$ and $T_2$, for an oscillation period $T_{osc} \approx 400$~s. The experimental data (dots) are compared with the theoretical expectations (smooth lines).}
    \label{fig:concatenated_equal_memristors_time}
\end{figure}

In Suppl. Figures \ref{fig:cicli}-\ref{fig:concatenated_equal_memristors_time} we report the hysteresis loops that characterize the effect of a single and two cascaded memristors, at different configurations of the integration times.

\newpage
    
\section{Simulation with different input functions}
\begin{figure}[h!]
    \centering
    \includegraphics[width=0.8\textwidth]{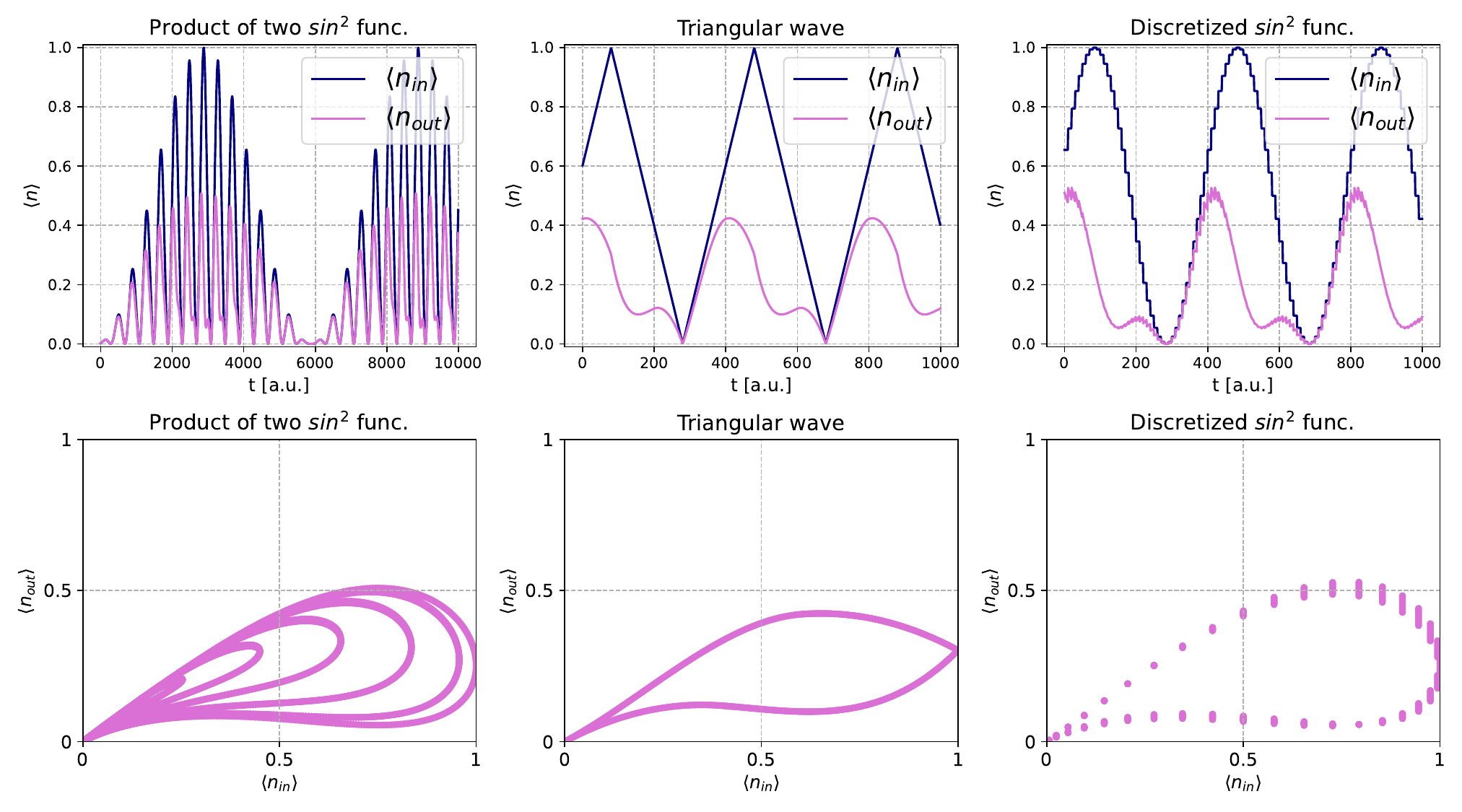}
    \caption{\textbf{Dynamics of a memristor with different input signals.} We show the simulated dynamics of a memristor taking four different input signals. In the upper row the temporal shape are reported, while in the bottom row there are the correspondent hysteresis loops.}
    \label{fig:change_input}
\end{figure}

In Suppl. Fig.~\ref{fig:change_input} we report the simulated memristive dynamics when the one-photon population of the input state oscillate differently than the square sine.

\newpage

\section{Study of the memristive dynamics of $N$ concatenated memristors}

\begin{figure}[h!]
    \centering
    \includegraphics[width=0.8\textwidth]{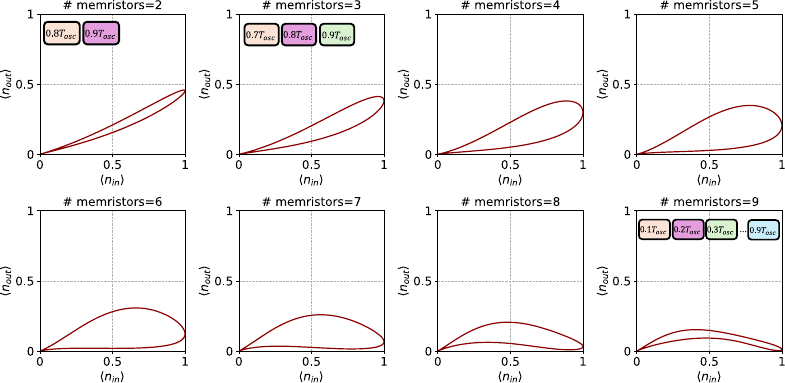}
    \caption{\textbf{Dynamics of variable number of concatenated memristors with increasing integration times in reverse.} We show the simulated dynamics of a chain of concatenated memristors with increasing time integration fractions with respect to the period of the input signal. For example, in the first plot with two memristors, the first memristor has a characteristic integration time $T = 0.8 T_{osc}$ and the second $T = 0.9 T_{osc}$ while, in the last plot with nine memristors, the first memristor has  $T = 0.1 T_{osc}$, the second $T = 0.2 T_{osc}$, and so on, until the last one having $T = 0.9 T_{osc}$. }
    \label{fig:increasing_fractions_from_0_9}
\end{figure}

In Suppl. Fig.~\ref{fig:increasing_fractions_from_0_9} we analyze through numerical simulations the evolution of the hysteresis loop of a chain of $N$ memristors when adding them progressively, with integration times varying from two to nine memristors. 
Considering that the area of a memristive hysteresis loop is linked to its memory capacity \cite{pfeiffer2016quantum}, simulations of this kind can be helpful to maximize this quantity by suitably varying the QM characteristic time.

\newpage
